\documentclass[english,final]{revtex4}
\usepackage[T1]{fontenc}
\usepackage[latin9]{inputenc}
\usepackage{amsmath}
\usepackage{graphicx}
\usepackage{esint}

\makeatletter
\@ifundefined{textcolor}{}
{%
 \definecolor{BLACK}{gray}{0}
 \definecolor{WHITE}{gray}{1}
 \definecolor{RED}{rgb}{1,0,0}
 \definecolor{GREEN}{rgb}{0,1,0}
 \definecolor{BLUE}{rgb}{0,0,1}
 \definecolor{CYAN}{cmyk}{1,0,0,0}
 \definecolor{MAGENTA}{cmyk}{0,1,0,0}
 \definecolor{YELLOW}{cmyk}{0,0,1,0}
}

\makeatother

\usepackage{babel}
\begin{document}
\title{Dispersive Two-Loop Calculations: Methodology and Applications}
\author{A. Aleksejevs, S. Barkanova}
\affiliation{Grenfell campus of Memorial University, Corner Brook, NL, Canada}
\begin{abstract}
As the new-generation precision experiments such as MOLLER \cite{MOLLER}
and P2 \cite{P2} look for physics beyond Standard Model, it is becoming
increasingly important to evaluate the higher-order electroweak radiative
corrections to a sub-percent level of uncertainty. However, due to
propagators with different masses and higher-order tensor Feynman
integrals, the two-loop calculations involving thousands of Feynman
graphs become a demanding task requiring novel computational approaches.
In this paper, we describe our dispersive sub-loop insertion approach
and develop two-loop integrals using two-point functions basis which
is applicable to wide range of processes.
\end{abstract}
\maketitle

\section{Introduction}

In the past decade, the search for physics beyond the Standard Model
(BSM) became one of the most important objectives in particle
physics. The searches for BSM physics involve high-energy colliders, underground,
ground and space telescopes, and high-precision experiments with 
high intensity beams at low energies. With high-precision
searches, the measured observables, such as left-right (LR) or forward-backward
(FB) parity-violating asymmetries, are extracted with uncertainties
reaching a percent level. Any significant deviation between theoretical
prediction based on the Standard Model (SM) calculation and experiment would be a definitive
signal for BSM physics. The MOLLER experiment planned at JLab  \cite{MOLLER} is proposing to measure
the PV asymmetry in the electron-electron scattering with the fractional accuracy
of 2.4\%, which is more than a factor of five improvement over the precision
of its predecessor experiment E-158 at SLAC \cite{E-158}. The P2 experiment  \cite{P2}  proposes to measure
PV asymmetry in electron-proton scattering with overall fractional accuracy at 1.4\%. Obviously,
the theoretical uncertainty must be lower or at least match the experimental
accuracy to make any conclusions regarding the BSM physics signal. The theoretical
accuracy is mainly derived from propagation of uncertainty in input
parameters and from limited knowledge of the higher order, i.e. beyond the one-loop
level, contributions. Specifically, for the MOLLER experiment, the major electroweak two-loop corrections to the Born asymmetry evaluated in \cite{Q-Part,T-part-1,T-part-2,2loopBox,2-loop-vertex} in the on-shell renormalization scheme, were found to be close to five percent, which is a significant contribution compared to the expected experimental precision. Clearly,
it is imperative to calculate a full set of two-loop diagrams participating
in $e-e$ or $e-p$ PV scattering, but this is not a straightforward
task, and it will most probably require a high degree of automatization due to a very large number of diagrams.
There is an extensive body of literature dedicated to the development of two-loop calculations  \cite{Kreimer,Czarnecki,Frink,Adams1,Adams2,Adams3,Remiddi1,Bloch1,Bloch2,Bohm,Hollik-1,Hollik-2,Gluza2005,Gluza2008,Freitas},
 offering a wide spectrum of approaches. We have outlined our general approach to calculations of the two-loop
diagrams based on the representation of many-point Passarino-Veltman
(PV) functions in two-point function basis in \cite{AAPRD2018,AAX2019,AADimeReg2019}. Here, we were
able to replace a sub-loop integral by the dispersive and regularized
representation of the two-point function. As a consequence, the second-loop integral received an additional propagator and we were able to
use the PV basis for the second-loop integration in the final stage
of the calculations. In this paper, we outline some of the results obtained with the approach developed in \cite{AAPRD2018,AAX2019,AADimeReg2019}.

\section{Sub-Loop Insertion}

We derive the main ideas in the dispersive sub-loop approach from
the example of the self-energy and triangle insertions.
\begin{figure}
\begin{centering}
\includegraphics[scale=0.6]{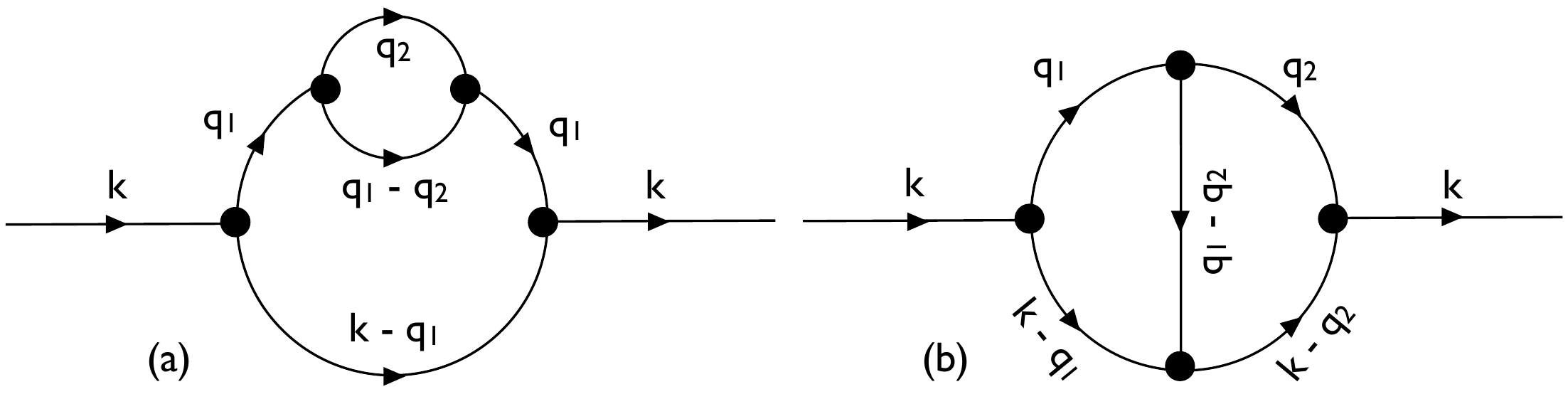}
\par\end{centering}
\caption{Two-loop self-energy and triangle insertions.}

\label{figSE2-TR2}
\end{figure}
For the the left graph in Fig. \ref{figSE2-TR2} , we can write the
following two-loop integral:
\begin{align}
M_{a}= & -\frac{2}{\pi^{4}}\intop\frac{d^{4}q_{1}d^{4}q_{2}}{\left[q_{2}^{2}-m^{2}\right]\left[\left(q_{1}-q_{2}\right)^{2}-m^{2}\right]\left[q_{1}^{2}-m^{2}\right]^{2}\left[\left(k-q_{1}\right)^{2}-m^{2}\right]}.\label{eq1}
\end{align}
Here, we assume that all propagators represent scalar particles,
couplings are set to one, and masses are the same. Integration over
sub-loop momentum $q_{2}$ will result in the simple two-point function:
$B_{0}\left(q_{1}^{2},m^{2},m^{2}\right)=-\frac{i}{\pi^{2}}\intop\frac{d^{4}q_{2}}{\left[q_{2}^{2}-m^{2}\right]\left[\left(q_{1}-q_{2}\right)^{2}-m^{2}\right]}$:
\begin{align}
M_{a}=- & \frac{2i}{\pi^{2}}\lim_{\phi\rightarrow0}\frac{\partial}{\partial\phi}\intop\frac{B_{0}\left(q_{1}^{2},m^{2},m^{2}\right)d^{4}q_{1}}{\left[q_{1}^{2}-\left(m^{2}+\phi\right)\right]\left[\left(k-q_{1}\right)^{2}-m^{2}\right]}.\label{eq1a}
\end{align}
To keep the results in the two-point function basis, we have removed quadratic
form $1/\left(q_{1}^{2}-m^{2}\right)^{2}$ and replaced it by $\lim_{\phi\rightarrow0}\frac{\partial}{\partial\phi}\left(1/\left[q_{1}^{2}-\left(m^{2}+\phi\right)\right]\right)$.
Using \cite{AADimeReg2019}, we can write the two-point insertion $B_{0}\left(q_{1}^{2},m^{2},m^{2}\right)$
dispersively. The sub-loop insertion does not include terms linear
in $\epsilon=\frac{4-D}{2}$ since the second loop integral is UV-finite. Applying
$\overline{MS}$ subtraction at the scale $\Lambda$, we can write
the sub-loop insertion as follows:
\begin{align}
B_{0}\left(q_{1}^{2},m^{2},m^{2}\right) & =\ln\frac{\Lambda^{2}}{m^{2}}-\frac{q_{1}^{2}}{\pi}\intop_{4m^{2}}^{\infty}ds\frac{\Im B_{0}\left(s,m^{2},m^{2}\right)}{s\left[q_{1}^{2}-s-i\varepsilon\right]}.\label{eq2}
\end{align}
Eq.\ref{eq2} will contribute an additional propagator to the
second loop, thus using Eq.\ref{eq2} in Eq.\ref{eq1a}, we can produce
the following two-loop result:
\begin{align}
M_{a}= & 2\ln\frac{\Lambda^{2}}{m^{2}}\partial_{\phi}B_{0}\left(k^{2},m^{2},m^{2}+\phi\right)\mid_{\phi=0}\nonumber \\
\nonumber \\
 & -\frac{2}{\pi}\intop_{4m^{2}}^{\infty}ds\frac{\Im B_{0}\left(s,m^{2},m^{2}\right)}{s\left(s-m^{2}\right)^{2}}\biggl[m^{2}\left(m^{2}-s\right)\partial_{\phi}B_{0}\left(k^{2},m^{2},m^{2}+\phi\right)\mid_{\phi=0}\label{eq:3}\\
\nonumber \\
 & +s\left(B_{0}\left(k^{2},m^{2},s\right)-B_{0}\left(k^{2},m^{2},m^{2}\right)\right)\biggr].\nonumber 
\end{align}
The dispersive integrand in Eq.\ref{eq:3} does not have any $1/\epsilon$
poles, thus we can neglect all terms containing $\epsilon$ dependence.
Here, two-point scalar function, its derivative and imaginary part
have the simple analytical structure:
\begin{align}
B_{0}\left(k^{2},m_{1}^{2},m_{2}^{2}\right)= & 2+\ln\frac{\mu^{2}}{m_{2}^{2}}+\frac{\sqrt{\lambda}}{k^{2}}\ln\frac{\Delta_{+}+\sqrt{\lambda}}{2m_{1}m_{2}}-\frac{\Delta_{-}}{2k^{2}}\ln\frac{m_{1}^{2}}{m_{2}^{2}},\nonumber \\
\label{eq:4}\\
\partial_{\phi}B_{0}\left(k^{2},m^{2},m^{2}+\phi\right)\mid_{\phi=0}= & -\frac{1}{\sqrt{\Delta_{0}k^{2}}}\ln\frac{\sqrt{\Delta_{0}k^{2}}-k^{2}+2m^{2}}{2m^{2}},\nonumber \\
\nonumber \\
\Im B_{0}\left(s,m^{2},m^{2}\right)= & \pi\sqrt{1-\frac{4m^{2}}{s}},\nonumber 
\end{align}
with $\Delta_{\pm}=m_{1}^{2}\pm m_{2}^{2}\mp k^{2}$, $\Delta_{0}=k^{2}-4m^{2}$,
and $\lambda\equiv\lambda\left(k^{2},m_{1}^{2},m_{2}^{2}\right)=k^{4}+m_{1}^{4}+m_{2}^{4}-2\left(k^{2}m_{1}^{2}+k^{2}m_{2}^{2}+m_{1}^{2}m_{2}^{2}\right)$
is a usual Kallen function. In the final steps of calculations, the dispersive
integration in the Eq.\ref{eq:3} can be done numerically.

For the right graph in the Fig.\ref{figSE2-TR2}, we can write the two-loop
integral as:
\begin{align}
M_{b}= & -\frac{1}{\pi^{4}}\intop\frac{d^{4}q_{1}d^{4}q_{2}}{\left[q_{2}^{2}-m^{2}\right]\left[\left(k-q_{2}\right)^{2}-m^{2}\right]\left[\left(q_{1}-q_{2}\right)^{2}-m^{2}\right]\left[q_{1}^{2}-m^{2}\right]\left[\left(k-q_{1}\right)^{2}-m^{2}\right]}.\label{eq:6}
\end{align}
Clearly, the integration over momentum $q_{2}$ is represented by the
three-point Passarino-Veltman function $C_{0}^{\left\{ 1\right\} }\equiv C_{0}\left(k^{2},\left(k-q_{1}\right)^{2},q_{1}^{2},m^{2},m^{2},m^{2}\right)=-\frac{i}{\pi^{2}}\intop\frac{d^{4}q_{2}}{\left[q_{2}^{2}-m^{2}\right]\left[\left(k-q_{2}\right)^{2}-m^{2}\right]\left[\left(q_{1}-q_{2}\right)^{2}-m^{2}\right]}$.
In order to replace $C_{0}^{\left\{ 1\right\} }$ function by the
propagator-like structure, we need to write a dispersive
representation of the three-point function. Using ideas from \cite{AAPRD2018,AADimeReg2019},
we can use the Feynman trick to join the first two propagators in
Eq.\ref{eq:6}, remove the quadratic form, and, after shifting momentum
$q_{2}-q_{1}=\tau$, write $C_{0}^{\left\{ 1\right\} }$ function
as:
\begin{align}
C_{0}^{\left\{ 1\right\} }= & -\frac{i}{\pi^{2}}\lim_{\phi\rightarrow0}\frac{\partial}{\partial\phi}\intop_{0}^{1}dx\intop\frac{d^{4}\tau}{\left[\tau^{2}-m^{2}\right]\left[\left(\tau-\left(kx-q_{1}\right)\right)^{2}-\left(m_{12}^{2}+\phi\right)\right]}\nonumber \\
\label{eq:7}\\
= & \lim_{\phi\rightarrow0}\frac{\partial}{\partial\phi}\intop_{0}^{1}dxB_{0}\left(\left(kx-q_{1}\right)^{2},m^{2},\left(m_{12}^{2}+\phi\right)\right),\nonumber 
\end{align}
where $m_{12}^{2}=m^{2}-k^{2}\bar{x}x$, and $\bar{x}$ is defined
as $\bar{x}=1-x$. In the case when $k^{2}>4m^{2}$, the mass parameter
$m_{12}$ in the two-point function becomes imaginary for the values
of $x\in\left(x_{1},x_{2}\right)$, where $\{x_{1},x_{2}\}$ are
the real parts of roots of the equation $m^{2}-k^{2}\bar{x}x=0$.
As a result, it is required to modify the dispersive representation
of the two-point function. We provide the detailed discussion for the case of dispersive treatment of two-point functions
with imaginary masses in \cite{AAPRD2018}. Using \cite{AAPRD2018} and \cite{AADimeReg2019},
we can write the following:
\begin{align}
 & C_{0}^{\left\{ 1\right\} }=\intop_{0}^{1}dx\left[\frac{1}{k^{2}\bar{x}x}\left(1+\frac{m^{2}}{k^{2}\bar{x}x}\ln\frac{m_{12}^{2}}{m^{2}}\right)+\left(kx-q_{1}\right)^{2}F\left(\left(kx-q_{1}\right)^{2},m^{2},m_{12}^{2}\right)\right],\nonumber \\
\nonumber \\
 & F\left(\left(kx-q_{1}\right)^{2},m^{2},m_{12}^{2}\right)=\begin{cases}
\frac{1}{\pi}\intop_{\left(m+m_{12}\right)^{2}}^{\infty}ds\frac{\Im\left[\partial_{\phi}B_{0}\left(s,m^{2},m_{12}^{2}+\phi\right)\mid_{\phi=0}\right]}{\left[s-i\omega\right]\left[s-\left(kx-q_{1}\right)^{2}-i\omega\right]}, & \textrm{if}\;x\in\left[0,x_{1}\right]\cup\left[x_{2},1\right]\\
\\
\frac{1}{2\pi i}\intop_{-\infty}^{\infty}ds\frac{\partial_{\phi}B_{0}\left(s,m^{2},m_{12}^{2}+\phi\right)\mid_{\phi=0}}{\left[s-i\omega\right]\left[s-\left(kx-q_{1}\right)^{2}-i\omega\right]} & \textrm{if}\;x\in\left(x_{1},x_{2}\right)
\end{cases}\label{eq:8}
\end{align}
and 
\begin{align*}
\partial_{\phi}B_{0}\left(s,m^{2},m_{12}^{2}+\phi\right)\mid_{\phi=0} & =\frac{1}{2s}\ln\frac{m^{2}}{m_{12}^{2}}-\frac{m^{2}-m_{12}^{2}+s}{s\sqrt{\lambda\left(s,m^{2},m_{12}^{2}\right)}}\ln\frac{m^{2}+m_{12}^{2}-s+\sqrt{\lambda\left(s,m^{2},m_{12}^{2}\right)}}{2mm_{12}}.
\end{align*}
At this point, Eq.\ref{eq:6} can be expressed using the three-point dispersive
representation given in Eq.\ref{eq:8}:
\begin{align}
M_{b}= & -\frac{i}{\pi^{2}}\intop d^{4}q_{1}\frac{C_{0}^{\left\{ 1\right\} }}{\left[q_{1}^{2}-m^{2}\right]\left[\left(k-q_{1}\right)^{2}-m^{2}\right]}\nonumber \\
\nonumber \\
= & \intop_{0}^{1}\frac{dx}{k^{2}\bar{x}x}\left(1+\frac{m^{2}}{k^{2}\bar{x}x}\ln\frac{m_{12}^{2}}{m^{2}}\right)B_{0}^{\left\{ 1\right\} }\label{eq:10}\\
\nonumber \\
 & -\frac{1}{\pi}\left(\intop_{0}^{x_{1}}dx+\intop_{x_{2}}^{1}dx\right)\intop_{\left(m+m_{12}\right)^{2}}^{\infty}ds\frac{\Im\left[\partial_{\phi}B_{0}\left(s,m^{2},m_{12}^{2}+\phi\right)\mid_{\phi=0}\right]}{s-i\omega}\left(B_{0}^{\left\{ 1\right\} }+sC_{0}^{\left\{ 2\right\} }\right)\nonumber \\
\nonumber \\
 & -\frac{1}{2\pi i}\intop_{x_{1}}^{x_{2}}dx\intop_{-\infty}^{\infty}ds\frac{\partial_{\phi}B_{0}\left(s,m^{2},m_{12}^{2}+\phi\right)\mid_{\phi=0}}{s-i\omega}\left(B_{0}^{\left\{ 1\right\} }+sC_{0}^{\left\{ 2\right\} }\right),\nonumber 
\end{align}
where $B_{0}^{\left\{ 1\right\} }\equiv B_{0}\left(k^{2},m^{2},m^{2}\right)$
and $C_{0}^{\left\{ 2\right\} }\equiv C_{0}\left(k^{2},k^{2}\bar{x}^{2},k^{2}x^{2},m^{2},m^{2},s\right)$.
In Eq.\ref{eq:10}, the two-point function $B_{0}^{\left\{ 1\right\} }$
is UV divergent, but $M_{b}$ (right graph on Fig.\ref{figSE2-TR2})
is UV-finite and hence should not contain any dependence on $1/\epsilon^{n}$
or scale parameter $\mu$. This dependence cancels out when we calculate
$M_{b}$ numerically, which provides a good test of Eq.\ref{eq:10}.
\begin{figure}
\begin{centering}
\includegraphics[scale=0.29]{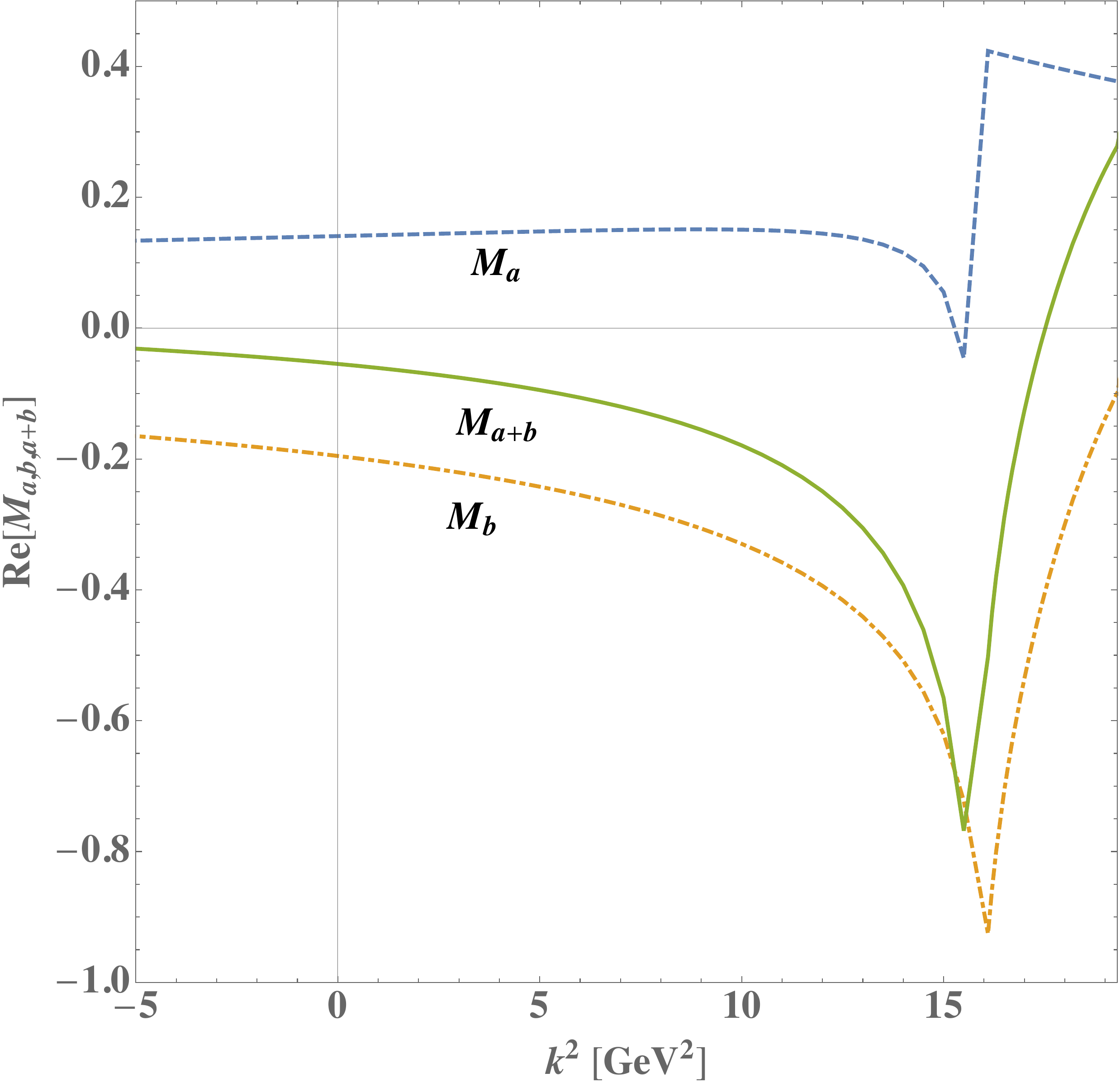} \includegraphics[scale=0.295]{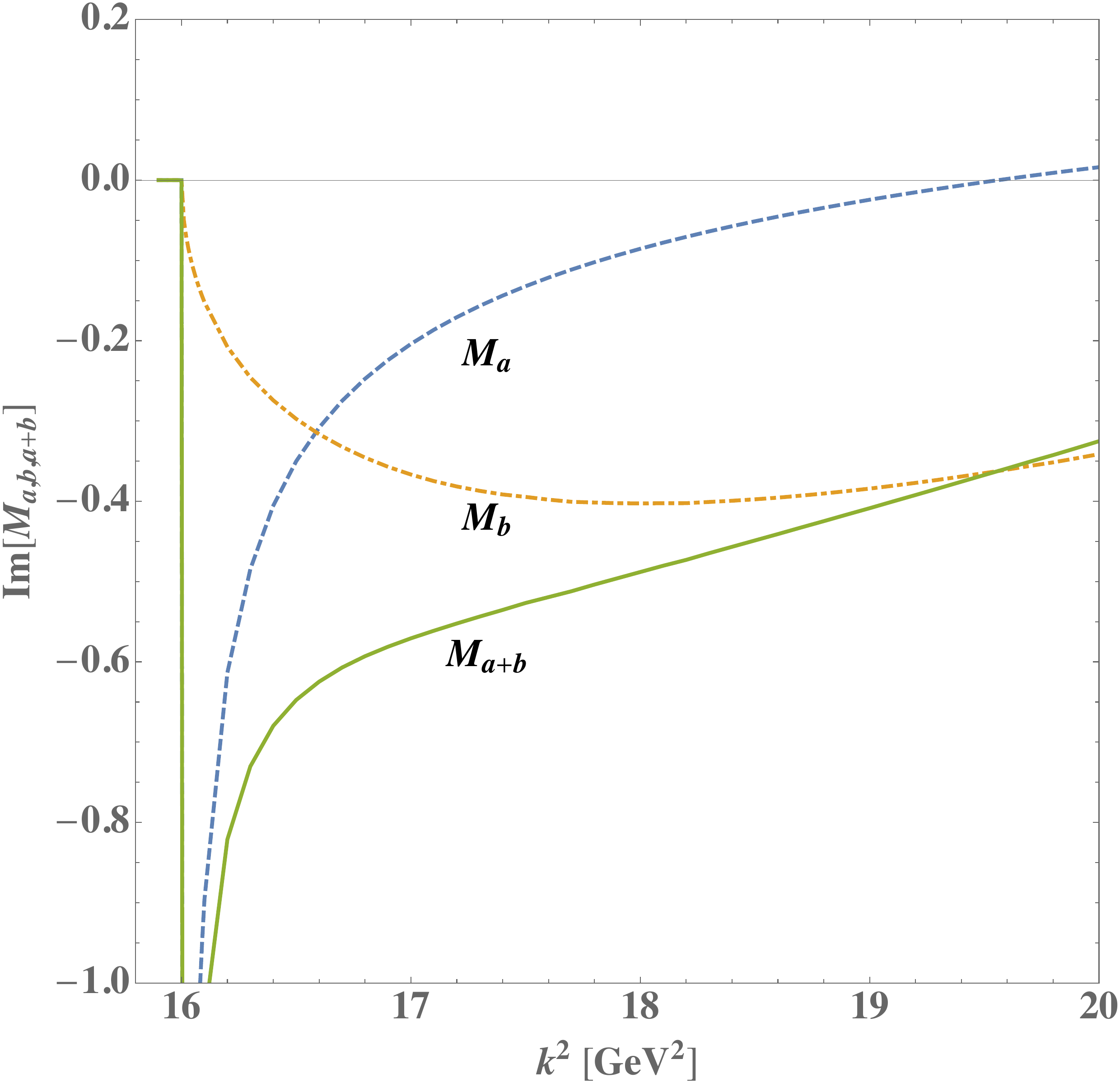}
\par\end{centering}
\caption{The left plot shows dependence of the real parts of Feynman diagrams
shown on Fig.\ref{figSE2-TR2} as a function of four-momentum squared
$k^{2}$ for the $m=2.0$ (GeV). The right plot gives the dependence
of imaginary parts of the same diagrams for the above threshold conditions
$k^{2}>4m^{2}$. Scale parameter $\Lambda$ is set to 2.0 (GeV).}

\label{fig2-results}
\end{figure}
For both dispersive and Feynman parameter numerical integration, we
use Gauss-Kronrod integration library. In order to keep Feynman parameter
integration stable, we have added a small imaginary part to the mass
$m$. The results shown on Fig.\ref{fig2-results} are given for both
real and imaginary parts below and above the threshold conditions. On
the left plot (see Fig.\ref{fig2-results}), contributions from $\Re\left[M_{a}\right]$
and $\Re\left[M_{b}\right]$ into $\Re\left[M_{a+b}\right]$ do show
some degree of cancellation in the space-like regime, and resonance
type behaviour near the threshold. The computing time of the dispersion
integral is in the order of fraction of a second. As for Feynman parameter
integration, computing time highly depends on the threshold and is usually in the order of a few seconds
per point below the threshold. Above the threshold, computing time raises dramatically (few
minutes per point) due to the numerical noise at the points $x_{1}$
and $x_{2}$. Overall, Eq.\ref{eq:3} and Eq.\ref{eq:10}, are in compact
form and applicable for the broad kinematic region.

\section{Conclusion}

In this paper, we have outlined the dispersive treatment approach of
the sub-loop insertion and represented the two-loop results in the
two-point function basis. The second-loop integration was reduced
to the one-loop type Feynman graph with an additional propagator coming
from the dispersive sub-loop insertion. As an example, we chose the two-loop scalar self-energy calculations and the corresponding  numerical results in Fig.\ref{fig2-results}. The main conclusion of this work is that dispersive insertion approach simplifies
analytical expressions considerably, to the point that it is possible
to employ computer algebra evaluating the two-loop calculations analytically
and carry out integration numerically.

\begin{acknowledgments}
The authors are grateful to the Lepton Photon 2019 organizing committee. This work is supported by the Natural Sciences and Engineering Research Council of Canada.
\end{acknowledgments}

\end{document}